\documentclass[sigconf]{acmart-me}

\AtBeginDocument{%
  \providecommand\BibTeX{{%
    \normalfont B\kern-0.5em{\scshape i\kern-0.25em b}\kern-0.8em\TeX}}}

\usepackage{amsmath,amsfonts}
\usepackage{algorithmic}
\usepackage{graphicx}
\usepackage{textcomp}
\usepackage{xcolor}
\usepackage{booktabs}
\usepackage{url}
\usepackage{multirow}
\usepackage{enumitem}
\usepackage{array, makecell}
\usepackage{hyperref}
\usepackage{float}
\hypersetup{colorlinks,allcolors=black}

\DeclareMathOperator{\E}{\mathbb{E}}

\usepackage[acronym, nowarn]{glossaries}
\makeglossaries

\newacronym{gi}{GI}{gastrointestinal}
\newacronym{ai}{AI}{artificial intelligence}
\newacronym{cadx}{CADx}{computer-aided diagnosis}
\newacronym{ml}{ML}{machine learning}
\newacronym{dl}{DL}{deep learning}
\newacronym{cnn}{CNN}{convolutional neural network}
\newacronym{crc}{CRC}{colorectal cancer}

\acmConference[MediaEval'20]{Multimedia Evaluation Workshop}{December 14-15 2020}{Online}
\setcopyright{rightsretained}
\copyrightyear{2020}
\acmYear{2020}
\acmDOI{}
\acmPrice{}
\acmISBN{}

\begin{document}
\title{Generative Adversarial Networks for Automatic Polyp Segmentation}

\author{Awadelrahman M. A. Ahmed}
\affiliation{%
 	 University of Oslo, Norway;  aahmed@ifi.uio.no 
}

\renewcommand{\shortauthors}{A. Ahmed}
\renewcommand{\shorttitle}{Generative Adversarial Networks for Automatic Polyp Segmentation}

\begin{abstract}
This paper aims to contribute in bench-marking the automatic polyp segmentation problem using generative adversarial networks framework. Perceiving the problem as an image-to-image translation task, conditional generative adversarial networks are utilized to generate masks conditioned by the images as inputs. Both generator and discriminator are convolution neural networks based. The model achieved 0.4382 on Jaccard index and 0.611 as F2 score.

\end{abstract}

\maketitle
\section{Introduction}\label{introduction}
Developing an automated computer-aided diagnosis system for polyp segmentation is indeed one of the potential solutions that can assist colonoscopy, the current gold-standard medical procedure for examining the colon, in the sense that can lessen the percentage of the overlooked polyps. Motivated by that, this paper contributes by examining a model based on generative adversarial networks (GANs) to represent one of the benchmark methods that other cutting edge new methods can be compared to. This paper discusses a submission to \emph{Medico automatic polyp segmentation challenge} \cite{Debesh} for task 1 which asks the participants to develop algorithms for segmenting polyps on a comprehensive data set. In this work we perceive the polyp segmentation problem as an image-to-image translation problem, as we are given a gastrointestinal  polyp image and the task is to translate it to the corresponding mask that locates the polyps. GANs framework have been successfully implemented to solve image-to-image translation problem in many application fields. Authors in \cite{isola2017image} investigated conditional GANs as a general-purpose solution to image-to-image translation problems and evaluated in different application fields such as translating aerial images to maps and reconstructing objects from edge maps or templates. The challenge task can be seen in the same way, hence we adapted the model architecture suggested in \cite{isola2017image} to fit the polyp segmentation problem. The following sections will illustrate the model details, evaluation and results followed by the conclusion and future work.

%
%\begin{figure} [!t]
%    \centering
%    \includegraphics[height=1.6cm]{figures/10_image.jpg}
%    \includegraphics[height=1.6cm]{figures/3_image.jpg}
%    \includegraphics[height=1.6cm]{figures/26.jpg}
%    \includegraphics[height=1.6cm]{figures/11image.jpg}
%    \includegraphics[height=1.6cm]{figures/3mask.jpg}
%    \includegraphics[height=1.6cm]{figures/10_mask.jpg}
%    \includegraphics[height=1.6cm]{figures/26mask.jpg}
%    \includegraphics[height=1.6cm]{figures/11mask.jpg}
%    \caption{Polyps and corresponding masks from Kvasir-SEG}  
%    \label{fig:images}
%\end{figure}
%
%
%
%\begin{figure} [!t]
%    \centering
%    \includegraphics[height=1.6cm]{figures/test_images/1.jpg}
%    \includegraphics[height=1.6cm]{figures/test_images/2.jpg}
%    \includegraphics[height=1.6cm]{figures/test_images/3.jpg}
%    \includegraphics[height=1.6cm]{figures/test_images/4.jpg}
%    \includegraphics[height=1.6cm]{figures/test_images/5.jpg}
%    \includegraphics[height=1.6cm]{figures/test_images/6.jpg}
%    \includegraphics[height=1.6cm]{figures/test_images/7.jpg}
%    \includegraphics[height=1.6cm]{figures/test_images/8.jpg}
%    \caption{Examples polyps from the test images}  
%    \label{fig:testimages}
%\end{figure}

\section{GANs for Polyp Segmentation}

The GANs framework is a generative model scheme based on a game-theoretic formulation for training data-synthesis models \cite{goodfellow2014generative}. A GAN consists of two models (e.g., neural networks) trained in opposition to each other. A generator network $G$ takes a random noise vector as an input and maps it to an output which represents fake data. The other network is the discriminator $D$, which receives that fake data and classifies it as fake data (i.e., generated by $G$) or as real data from a real dataset. Nevertheless, our particular task is to map each gastrointestinal polyp image to the corresponding mask that defines the polyps area, this description does not fit within the standard GAN setting described above where the input is a noise and the output is an arbitrary (but realistic-looking) image. Therefore, for that we utilize a prominent variation of GANs which produces an output that is \textit{conditioned} by its input and hence called \textit{conditional} GAN. 

The  polyp segmentation GAN-based model consists of two networks. A generator takes the images as input and tries to produce realistic-looking masks conditioned by this input, and a discriminator which is basically a classifier which has the access to the ground truth masks and tries to classify whether the generated masks are real or not. To stabilize the training, the images are concatenated with the masks (generated or real) before being fed to the discriminator. The ultimate goal is to find the generator parameter set $\theta^*_G$ that satisfies the minimax optimization problem is shown in (\ref{eq1}). We particularly draw a batch of images $x$ and their corresponding masks $y$ from their distributions $X$ and $Y$ respectively. We feed $x$ to the generator $G$ to produce a fake mask set $G(x)$. We then concatenate both masks (real and fake) with the input images, this is represented by $C$. These concatenated pairs are then fed to the discriminator $D$. To learn these models, both networks are updated in an adversarial fashion based on the discriminator output; meaning that the discriminator aims to maximize the exact function that the generator aims to minimize. Through training time, the generator will be able to generated realistic-looking masks that are good enough to deceive the discriminator, i.e. it can classify as real.

\begin{equation}
	\begin{split}
		\theta^*_G=min_{\theta_G} max_{\theta_D} & \E_{x,y\sim X,Y}[log D(C(x,y)]\\
		&+\E_{x\sim X}[log(1-D(C(x,G(x))]\label{eq1}
	\end{split}
\end{equation}

\begin{figure} [htpb]
    \centering
    \includegraphics[scale=0.35]{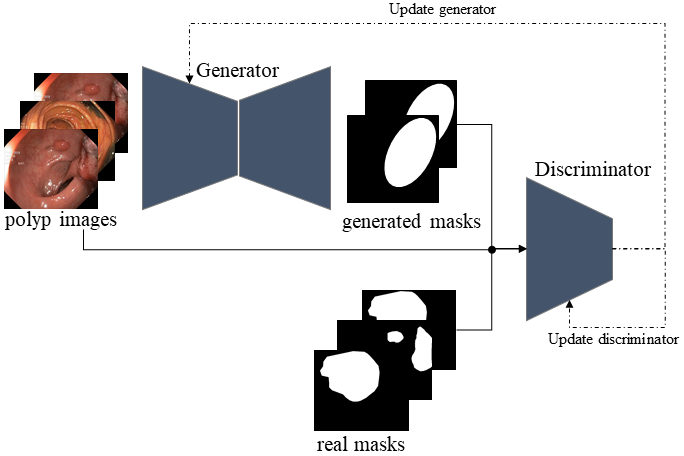}
    \caption{Model block diagram}  
    \label{bd}
\end{figure}

The model block diagram is shown in Figure \ref{bd} and the model details are shown in Table \ref{arch}. Both networks are based on convolution neural networks (CNNs). The generator has two segments, one is based on convolution operations folllowed by deconvolution layers with using skip connections.

%
%
%\begin{table}[htpb]
%	\centering
%	\begin{tabular}{|m{2cm}|m{4cm}|}
%		\hline
%		\multicolumn{2}{|c|}{\textit{Generator Network}}     \\
%		\hline
%		Conv1 &  $f$*1,\ \ \ \ LeakyReLU     \\
%  		\hline
% 		Conv2 &  $f$*2,\ \ \ \ LeakyReLU     \\
% 		\hline
%		Conv3 &  $f$*4,\ \ \ \ LeakyReLU     \\
%		\hline
%		Conv4 &  $f$*8,\ \ \ \ LeakyReLU     \\
%		\hline
%		Conv5 &  $f$*8,\ \ \ \ LeakyReLU     \\
%		\hline
%		Conv6 &  $f$*8,\ \ \ \ LeakyReLU     \\
%		\hline
%		Conv7 &  $f$*8,\ \ \ \ LeakyReLU     \\
%		\hline
%		Conv8 &  $f$*8,\ \ \ \ LeakyReLU     \\
%		\hline
%		DeConv1 &  $f$*8,\ \ \ \ ReLU,\ \ \ \ Dropout(0.5)    \\
%		\hline
%		DeConv2 &  $f$*8,\ \ \ \ ReLU,\ \ \ \ Dropout(0.5)     \\
%		\hline
%		DeConv3 &  $f$*8,\ \ \ \ ReLU,\ \ \ \ Dropout(0.5)     \\
%		\hline
%		DeConv4 &  $f$*8,\ \ \ \ ReLU,\ \ \ \ Dropout(0.5)     \\
%		\hline
%		DeConv5 &  $f$*4,\ \ \ \ ReLU,\ \ \ \ Dropout(0.5)     \\
%		\hline
%		DeConv6 &  $f$*2,\ \ \ \ ReLU,\ \ \ \ Dropout(0.5)     \\
%		\hline
%		DeConv7 &  $f$*1,\ \ \ \ ReLU,\ \ \ \ Dropout(0.5)     \\
%		\hline
%		DeConv8 &  1 \ \ ,\ \ \ \ Tanh     \\
%		\hline
%		\multicolumn{2}{|c|}{\textit{Discriminator Network}}   \\
%		\hline
%		Conv1 &  $f$*1,\ \ \ \ LeakyReLU     \\
%		\hline
%		Conv2 &  $f$*2,\ \ \ \ LeakyReLU     \\
%		\hline
%		Conv3 &  $f$*4,\ \ \ \ LeakyReLU     \\
%		\hline
%		Conv4 &  1 \ \ ,\ \ \ \ Sigmoid     \\
%		\hline
%		\multicolumn{2}{r}{  $f$: number of filters=64}   \\
%	\end{tabular}
%	\caption{Model details}
%	    \label{arch}
%\end{table}

\begin{table}[htpb]
	\centering
	\begin{tabular}{|m{2cm}|m{4cm}|}
		\hline
		\multicolumn{2}{|c|}{\textit{Generator Network}}     \\
		\hline
		Conv1 &  $f$*1,\ \ \ \ LeakyReLU     \\
		\hline
		Conv2 &  $f$*2,\ \ \ \ LeakyReLU     \\
		\hline
		Conv3 &  $f$*4,\ \ \ \ LeakyReLU     \\
		\hline
		Conv4, 5, 6, 7, 8 &  $f$*8,\ \ \ \ LeakyReLU     \\
		\hline
		DeConv1, 2, 3 ,4 &  $f$*8,\ \ \ \ ReLU,\ \ \ \ Dropout(0.5)    \\
		\hline
		DeConv5 &  $f$*4,\ \ \ \ ReLU,\ \ \ \ Dropout(0.5)     \\
		\hline
		DeConv6 &  $f$*2,\ \ \ \ ReLU,\ \ \ \ Dropout(0.5)     \\
		\hline
		DeConv7 &  $f$*1,\ \ \ \ ReLU,\ \ \ \ Dropout(0.5)     \\
		\hline
		DeConv8 &  1 \ \ ,\ \ \ \ Tanh     \\
		\hline
		\multicolumn{2}{|c|}{\textit{Discriminator Network}}   \\
		\hline
		Conv1 &  $f$*1,\ \ \ \ LeakyReLU     \\
		\hline
		Conv2 &  $f$*2,\ \ \ \ LeakyReLU     \\
		\hline
		Conv3 &  $f$*4,\ \ \ \ LeakyReLU     \\
		\hline
		Conv4 &  1 \ \ ,\ \ \ \ Sigmoid     \\
		\hline
		\multicolumn{2}{r}{  $f$: number of filters=64}   \\
	\end{tabular}
	\caption{Model details}
	\label{arch}
\end{table}

%\subsection{Model Description}

%\subsection{Implementation Detail}

\section{Experimental Evaluation}
As required by the challenge, the data set used is the Kvaris-SEG~\cite{jha2020kvasir} training data set which consists of 1000 image-mask pairs. We firstly split the dataset to 800 training set and 200 validation set to optimize the parameters. After we got satisfied with the model, we re-trained the model on the whole 1000 data pairs and used to produce the masks of a separate 160 test set that we submitted. The model is implemented using Python\cite{python} and PyTorch\cite{paszke2017automatic} framework on a 2$\times$14-core Intel/128Gib machine.

\subsection{Data Preparation and Training}\label{prep}
\begin{figure} [htpb]
	\centering
	\captionsetup{justification=centering}
	\includegraphics[width=\linewidth ]{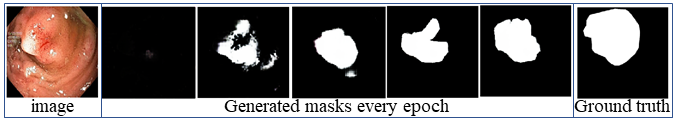}
	\caption{Generated masks in different epochs}  
	\label{epochs}
\end{figure}

As the images have different dimensions, we fit the images to the mean width and height by cropping the larger images and padding the smaller images with zeros. The pixel values are normalized between -1 and 1. No data augmentation were used. The networks in Table \ref{arch} were trained by optimizing the loss function in equation \ref{eq1}. The discriminator minimizes the negative log likelihood and the generator minimizes the negative of that. We tried other options like using feature matching, i.e. minimizing the loss between the features of the discriminator pre-last layer and the original masks instead of using the last classification layer output, but this seems to hurt the performance. We used Adam optimizer for both networks with learning rates of 0.002, for 12 epochs and batch size of 4. Some samples of the generated masks for some epochs during training is shown in Figure \ref{epochs}.

%\begin{figure} [htpb]
%	\centering
%	\includegraphics[width=\linewidth]{figures/losses.png}
%	\captionsetup{justification=centering}
%	
%	\caption{Generator (left) and discriminator (right) leaning curves}  
%	\label{losses}
%\end{figure}
%
%
%

\subsection{Results}\label{results}
To quantify the overlap percentage between the ground truth mask and our generated masks, we report both Jaccard index the Dice similarity coefficient (DSC). We also report the per pixel recall, precision, accuracy and F2 (giving more weight to recall). Results on the test set are in Table \ref{res}. However it is difficult to comment on those results objectively because of the uniqueness of the dataset \cite{jha2020kvasir}, i.e. no previous publications to compare with, but in general the higher recall than precision reflects the model accountability for false-negatives which is desirable for this application. The test throughput is 16 frames/sec on a 2$\times$14-core Intel/128Gib machine.
 
\begin{figure} [htpb]
	\centering
	\includegraphics[scale=0.6]{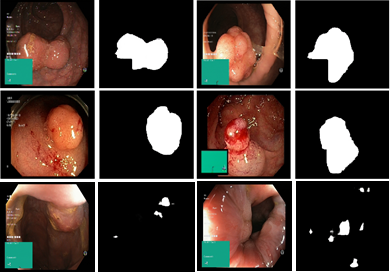}
	\caption{Model outputs from the test set}  
	\label{test}
\end{figure}

We report the model outputs for some samples in Figure \ref{test}. Even though we do not have the access to the ground truth of the test set, we may observe that the model incorrectly identified the polyp location of the bottom two samples; whereas it did far better in locating the polyps area of other samples (top and middle rows). We do not have a clear explanation for that, but we speculate that the small receptive field of the convolution layers could be a reason. In other words, the convolution layers pay attention to the close by area to each pixel and if this area is rich enough with features (e.g. has sharp edges) the polyp can be more distinguishable. This why incorporating attention mechanism \cite{zhang2019self} might help the model to attend to fine details and larger ranges; hence we suggest studying it as an extended model.

\begin{table}[htpb]
	\centering
	\begin{tabular}{|c|c|c|c|c|c|}
		\hline
		Jaccard& DSC & Recall & Precision & Accuracy &  F2    \\
		\hline
		0.4382 &  0.562 & 0.697 & 0.556 & 0.881 & 0.611   \\
		\hline
		
	\end{tabular}
	\caption{Test results as reported by the challenge organizers}
	\label{res}
\end{table}

%\subsection{Data Augmentation}

\section{Conclusion}\label{conclusion}
This paper aimed to benchmark the polyp segmentation problem using GANs. The problem has been perceived as an image-to-image translation task and we utilized conditional GANs architecture. The model was able to learn the masks however higher performance can be achieved by trying some improvements such as adding reconstruction loss and increasing the dataset with data augmentation. An interesting extension for this model is to try incorporating an attention layer \cite{zhang2019self} which can help the convolution layers in both generator and discriminator to attend to fine details and expand the receptive field.

\bibliographystyle{ACM-Reference-Format}
\bibliography{references}
\end{document}